\documentclass{cernrep}
\usepackage{texnames}
\usepackage[T1]{fontenc}
\newcommand{\ope}{$\omega_{pe}$}

\newcommand{\lpe}{$\lambda_{pe}$}
\newcommand{\kpe}{$k_{pe}$}

\newcommand{\neo}{$n_{e0}$}

\usepackage[colorlinks=true, linkcolor=black, citecolor=black, filecolor=black, urlcolor=blue]{hyperref}
\usepackage{fancyhdr}
\fancyhfoffset{4 mm}
\fancypagestyle{ARTTITLE}{%
\fancyhf{} 
\lhead{\footnotesize{Proceedings of the 2019 CERN--Accelerator--School course on
\it{ High Gradient Wakefield Accelerators}, Sesimbra, (Portugal)}}
\lfoot{Available online at \url{https://cas.web.cern.ch/previous-schools}}
\rfoot{\thepage\hspace*{3mm}}

   }
\begin{document}
\title{Beam-Driven Systems, Plasma Wakefield Acceleration}
 
\author{Patric Muggli}

\institute{Max Planck Institute for Physics, Munich, Germany}
 
\begin{abstract}
We expand on the material that was published in the previous Proceedings of the CERN Accelerator School on Plasma Wakefield Acceleration. %
The material focused on Plasma Wakefield Acceleration in the short, narrow bunch regime. %
After a brief introduction, we describe Plasma Wakefield Acceleration driven by a bunch train. %
We then attempt to give simple and intuitive descriptions of bunch self-modulation, occurring when the bunch is long, and of current filamentation, occurring when the bunch is wide. %
Self-modulation is a means to use existing bunches carrying large amounts of energy to drive plasma wakefields. %
Current filamentation instability imposes a limitation on how wide a particle bunch can be before transverse break-up may disrupt the~wakefields generation and the acceleration process. %
As in the previous material, we show sample experimental results that demonstrate that much of the physics at play has been observed experimentally. %
\end{abstract}
\maketitle 
\thispagestyle{ARTTITLE}
\section{Introduction}
The plasma wakefield accelerator (PWFA)~\cite{bib:chen} is one of the two plasma-based particle accelerators, the~other one being the laser wakefield accelerator (LWFA)~\cite{bib:tajima}. %
The process of driving wakefields is different between the LWFA, where the ponderomotive force of a laser pulse drives the wake in the initially neutral plasma, and the PWFA, where the transverse electric field of a relativistic particle bunch drives the wake. %
The acceleration process is identical in the two cases. %
In this text we therefore focus on the wakefields driving process. %

To some extent, PWFA physics is {\it simpler} than LWFA physics, mostly because the bunch that drives wakefields and the one that experiences them are of the same kind. %
Also, a particle bunch with parameters such that it drives wakefields similar to those driven by a laser pulse consists of fewer (of order 10$^{9}$ to 10$^{10}$), much higher energy particles ($\sim$MeV-GeV, electrons, protons, etc.) than does a laser pulse made of 10$^{16}$ to 10$^{18}$, $\sim$eV-energy photons. %
In particular, photons respond to the index of refraction of the plasma and to its changes. %
However, around the world there are many more laser systems capable of driving plasma wakefields than there are particle bunch sources. %
For years now, such lasers can be purchased {\it off the shelf}, whereas particle accelerators are typically found only at National Laboratories. %
Literature about LWFA is therefore also more abundant than that about PWFA. %
However, due to the many similarities, review articles often cover both. %
We strongly recommend reading general Refs.~\cite{bib:rev1} and~\cite{bib:rev2}. %
More specific articles on PWFA can be found in Refs.~\cite{bib:PWFA1} and~\cite{bib:PWFA2}. %

We build here on the text that appeared in the previous CAS Proceedings~\cite{bib:muggliCAS}. %
We also refer the~treader to the slides of the two lectures that contain additional figures that further illustrate some of the points made here. %
The text describes the fields of a relativistic charged particle and how it pertains to the driving of wakefields in plasma. %
After we described the expected field structure, we used a simple particle model following Ruth {\it et al.}~\cite{bib:ruth84} to show that the transformer ratio, defined as the ratio of peak accelerating field {\it behind} a single bunch E$_+$ and the peak decelerating field {\it within} the drive bunch E$_-$ is smaller or equal to two: R=E$_+$/E$_-\le$2. %
Here E$_+$  and E$_-$ are amplitudes and are therefore positive. %
This is know as the fundamental theorem of beam loading~\cite{bib:beamloadingtheo}. %

We then briefly introduced PWFA linear theory. %
We used it to show that for a single and short bunch, that is with length on the order of the wakefields wavelength or plasma skin depth, there is an~optimum length that depends on the bunch shape, that yields the largest transformer ratio. %
For a~Gaussian bunch with root-mean-square (RMS) length $\sigma_z$ this is expressed as: k$_{pe}\sigma_z\cong\sqrt{2}$. %
We also mentioned that the bunch transverse size must also be small, that is on the order or smaller than the plasma skin depth, which is usually expressed as k$_{pe}\sigma_r\le$1.\footnote{\kpe=\ope/c is the relativistic wakefields wave number in a plasma of density n$_{e0}$ and corresponding electron plasma angular frequency \ope=$\left(n_{e0}e^2/\epsilon_0m_e\right)^{1/2}$} %
We briefly touched on using multiple bunches to drive wakefields. %

We introduced what is usually considered as the order of magnitude for the maximum (amplitude of the) longitudinal (accelerating or decelerating) electric field that can be expected in a plasma: E$_{WB}$=$\frac{m_ec\omega_{pe}}{e}$. %

We also used linear theory to discuss beam loading by the witness bunch to be accelerated. %

We then jumped to the non-linear regime of the PWFA, in which one can use the linear variation of the ion column focusing field (and force on witness electrons) to preserve the incoming emittance of the~accelerating bunch. %
This is strictly the case either for each longitudinal slice of the bunch, or for a~bunch that remains mono-energetic, thanks for example to beam loading. %
That naturally led us to discuss the matching of the witness bunch to the ion column focusing force. %

We then showed a few experimental results that demonstrate that some of the effects presented~were actually observed in experiments: Variation of the longitudinal electric field along the drive bunch~\cite{muggli04}; Variation of the energy gain (and thus accelerating field) with relative drive bunch length, and plasma length and electron density~\cite{muggliNJP}; Preservation of the emittance of an electron bunch using betatron oscillations of the bunch envelope~\cite{muggli04,bib:betatron}; Non-preservation of the emittance of a positron bunch in a~plasma since the situation of the ideal pure ion column focusing for an electron does not exist for a~positron bunch~\cite{muggli08}. %

One of the major developments in the PWFA field is the results obtained in the AWAKE experiment~\cite{bib:muggli,bib:turner,bib:karl,bib:AWAKEgain} using a long proton bunch (k$_{pe}\sigma_z\gg$1). %
In the previous text!\cite{bib:muggliCAS} we mainly discussed PWFAs driven by small bunch drivers. By this we mean small when compared to the plasma wavelength \lpe=2$\pi$/\kpe or the plasma skin depth c/$\omega_{pe}$. %
We remind the reader that plasma electrons, that sustain the~wakefields by their collective response, oscillate at the plasma angular frequency $\omega_{pe}$. %
Since the~wakefields are {\it tied} to the driver, their phase velocity is equal to that of the driver with relativistic factor $\gamma_b$: v$_b$=$\left(1-1/\gamma_b^2\right)^{1/2}$c$\cong$c for an ultra-relativistic bunch with $\gamma_b\gg$1. %
This justifies k$_{pe}$=\ope/v$_b\cong\omega_{pe}$/c. %
We note here that these quantities are derived again in the context of dispersion relations, i.e., linear theory and small perturbations. %
That means for example that the plasma electrons velocity is {\it small}.\footnote{For example, in the context of an electro-static plasma wave, often used as the mode representing plasma wakefields, that means that first order quantities v$_{e1}$, n$_{e1}$ and E$_{z1}$ are {\it self-consistently} such that the electron plasma density perturbation is small when compared to the initial, uniform density n$_{e0}$: n$_{e1}\ll$n$_{e0}$ because the bunch density n$_b$ is small: n$_b\cong$n$_{e1}\ll$\neo. %
The two other quantities are zero in equilibrium: E$_{z0}$=0, v$_{e0}$=0.} %
However, wakefields can be driven into the non-linear regime. %
In this case the velocity of the plasma electrons can become relativistic, their mass increase m$_e\rightarrow\gamma_{pe}$m$_e$ ($\gamma_{pe}\ge$1) and thus $\omega_{pe}\rightarrow\omega_{pe}/\sqrt{\gamma_{pe}}$ decreases and $\lambda_{pe}\rightarrow\sqrt{\gamma_{pe}}\lambda_{pe}$ increases. %

Interestingly the history of the field of plasma-based acceleration started with experiments that used a long laser pulse to drive wakefields. %
Only long pulses with high enough intensity to drive large amplitude ($\sim$10 to 100\,MV/m) wakefields were available. %
Wakefields were driven either through laser pulse self-modulation~\cite{bib:SMlaser} or through beat-waves~\cite{bib:PBWA}. %
As soon as ultra-short, tens of femtoseconds long laser pulses became available, all experiments operated in the in the short pulse regime characterized by c$\tau_{laser}\le\lambda_{pe}/2$, $\tau_{laser}$ the laser pulse length. %
Nowadays there is a renewed interest in driving wakefields with multiple laser pulses, especially to reduce requirements for the laser system~\cite{bib:hooker}. %
PWFA history started with experiments at low gradients ($<$1\,GeV/m) because short, dense, relativistic electron bunches were not available. %
They operated already in the short bunch regime with k$_{pe}\sigma_z\cong$1~\cite{bib:rosenzweig}. %
We note that narrow bunches, \kpe$\sigma_r\ll$1 are interesting, especially in the non-linear regime, because wakefields parameters are then weakly dependent on the bunch radius and on its evolution, in particular because of betatron oscillations of the bunch envelope size (in an un-matched case, see for example~\cite{bib:VanDerMeer}). %

Only recently the idea of using self-modulation (SM) of a long particle bunch was proposed~\cite{bib:kumar}. %
This is mostly to take advantage of proton bunches available today and carrying large amounts of energy (tens to hundreds of kilojoules). %
Synchrotrons such as the CERN SPS and LHC produce these bunches routinely for high-energy and particle physics research. %
They are 6 to 12\,cm-long, but can be focused to small transverse sizes (e.g., 200\,$\mu$m~\cite{bib:muggli,bib:turner,bib:karl,bib:AWAKEgain}). %
The use of a single drive bunch in a single (or two) plasma(s) would avoid intricacies of staging~\cite{bib:staging} and would use existing drive bunches, i.e, no need to develop or build a new accelerator complex for the driver(s). %

We first use linear theory to build some intuition on driving wakefields with a train of bunches (or laser pulses). %
Then we discuss SM, a way to let the long bunch and plasma form the train through an instability, the self-modulation instability (SMI)~\cite{bib:kumar}, that can be controlled by seeded to turn it into a self-modulation (SSM) process~\cite{bib:muggli}. %
The plasma response that is the source of this longitudinal instability has an {\it equivalent} in the transverse plane, the current filamentation instability. %
We therefore discuss regimes in which either \kpe$\sigma_z\gg$1 or  \kpe$\sigma_r\gg$1. %

\section{Brief linear theory reminder}\label{sec:lintheo}

Wakefields for a bunch, i.e., a large collection of single particles such that the particle aspect can be neglected is calculated from Green's function, that is the wakefields driven by a single particle. %
Green's function for the longitudinal wakefields is derived in a 1D model for example in Ref.~\cite{bib:ruth85}. %
Wakefields can be calculated by convolving the bunch distribution with Green's function (see for example Ref.~\cite{bib:keinings} in 2D). %
The 2D expressions for cylindrically symmetric bunches are:
\begin{equation}\label{eq:MultiBunches_long1}
W_z (\xi, r)=-\frac{e}{\epsilon_{0}}\int_{-\infty}^{\xi} n_{b_{\parallel}}(\xi')cos[k_{pe}(\xi-\xi')]d\xi'\cdot R(r),
\end{equation}
\begin{equation}\label{eq:MultiBunches_tran1}
W_{\perp} (\xi, r)=\frac{e}{\epsilon_{0}k_{pe}}\int_{-\infty}^{\xi} n_{b_{\parallel}}(\xi')sin[k_{pe}(\xi-\xi')]d\xi'\cdot \frac{dR(r)}{dr}, 
\end{equation}
where $R(r)$ is the transverse dependency given by:
\begin{equation}\label{eq:MultiBunches_Rr}
\begin{array}{rcl}
R(r)={k_{pe}}^2\int_0^rr'dr'n_{b\perp}(r')I_0(k_{pe}r')K_0(k_{pe}r)+{k_{pe}}^2\int_r^\infty r'dr'n_{b\perp}(r')I_0(k_{pe}r)K_0(k_{pe}r'),
 \end {array} 
\end{equation}
and $I_0$ and $K_0$ are the zeroth order modified Bessel functions of the first and second kind, respectively. %

For a bunch with constant density over $0\le\xi\le\xi_b$ and within that $\xi$ range:
\begin{equation}\label{eq:MultiBunches_long2}
W_z (0\le\xi\le\xi_b)\propto n_{b0}\int_{0}^{\xi}cos[k_{pe}(\xi-\xi')]d\xi',
\end{equation}
and
\begin{equation}\label{eq:MultiBunches_tran2}
W_{\perp} (0\le\xi\le\xi_b)\propto n_{b0}\int_{0}^{\xi}sin[k_{pe}(\xi-\xi')]d\xi'. 
\end{equation}
These equation can be integrated by writing $a=k_{pe}(\xi-\xi')$, thus $da=-k_{pe}\xi'$. %
Since $\xi\in[0,\xi]$, $a\in[k_{pe}\xi,0]$. %
Thus:
\begin{equation}\label{eq:MultiBunches_long3}
\begin{array}{lcl}
W_z (0\le\xi\le\xi_b)\propto n_{b0}\int_{k_{pe}\xi}^{0}cos[a]\left(\frac{-1}{k_{pe}}\right)da=\frac{n_{b0}}{k_{pe}}sin[a]|_{0}^{k_{pe}\xi}\\=\frac{n_{b0}}{k_{pe}}sin[k_{pe}\xi]
\end{array}
\end{equation}
and
\begin{equation}\label{eq:MultiBunches_tran3}
\begin{array}{lcl}
W_{\perp} (0\le\xi\le\xi_b)\propto n_{b0}\int_{k_{pe}\xi}^{0}sin[a]\left(\frac{-1}{k_{pe}}\right)da=\frac{n_{b0}}{k_{pe}}cos[a]|_{0}^{k_{pe}\xi}\\=\frac{n_{b0}}{k_{pe}}\left(1-cos[k_{pe}\xi]\right). 
\end{array}
\end{equation}
Equation~\ref{eq:MultiBunches_long3} shows that the longitudinal wakefields within the bunch are simply oscillating from its front and are alternatively (if the bunch is longer than half a plasma wavelength) decelerating (initially) and accelerating (can be shown from the sign and physics says so!). %
Since $-1\le cos[k_{pe}\xi]\le1$, Eq.~(\ref{eq:MultiBunches_tran3}) shows that the transverse wakefields do not change sign along the bunch. %
They are focusing for charges with the same sign as that of the drive bunch (can be shown from the sign and physics says so!). %

The same procedure can be used for the fields behind the drive bunch, but this time, since the~bunch charge exists only for $0\le\xi\le\xi_b$ one uses: $a=k_{pe}(\xi-\xi')$, thus $da=-k_{pe}\xi'$ and $\xi\in[0,\xi_b]$, $a\in[k_{pe}\xi,k_{pe}(\xi-\xi_b)]$. %
Thus:
\begin{equation}\label{eq:MultiBunches_long4}
\begin{array}{lcl}
W_z (\xi>\xi_b)\propto n_{b0}\int_{k_{pe}\xi}^{k_{pe}(\xi-\xi_b)}cos[a](\frac{-1}{k_{pe}})da=\frac{n_{b0}}{k_{pe}}sin[a]|_{k_{pe}\xi}^{k_{pe}(\xi-\xi_b)}\\
=\frac{n_{b0}}{k_{pe}}\left(sin[k_{pe}(\xi-\xi_b)]-sin[k_{pe}\xi]\right)
\end{array}
\end{equation}
and
\begin{equation}\label{eq:MultiBunches_tran4}
\begin{array}{lcl}
W_{\perp} (\xi>\xi_b)\propto n_{b0}\int_{k_{pe}\xi}^{k_{pe}(\xi-\xi_b)}sin[a](\frac{-1}{k_{pe}})da=\frac{n_{b0}}{k_{pe}}cos[a]|_{k_{pe}(\xi-\xi_b)}^{k_{pe}\xi}\\
=\frac{n_{b0}}{k_{pe}}\left(cos[k_{pe}(\xi-\xi_b)]-cos[k_{pe}\xi]\right). 
\end{array}
\end{equation}

Equations~\ref{eq:MultiBunches_long3},~\ref{eq:MultiBunches_tran3} and~\ref{eq:MultiBunches_long4},~\ref{eq:MultiBunches_tran4} can be plotted for any value of $\xi_b$. %
Similar calculations of linear theory wakefields for bunches with cosine shapes can be found in~\cite{bib:mathiaslin}. %
Cosine shape are useful in calculations and numerical simulations because of their final spatial extent (as opposed to Gaussian profile ones). %

\begin{figure}
\centering\includegraphics[width=.9\linewidth]{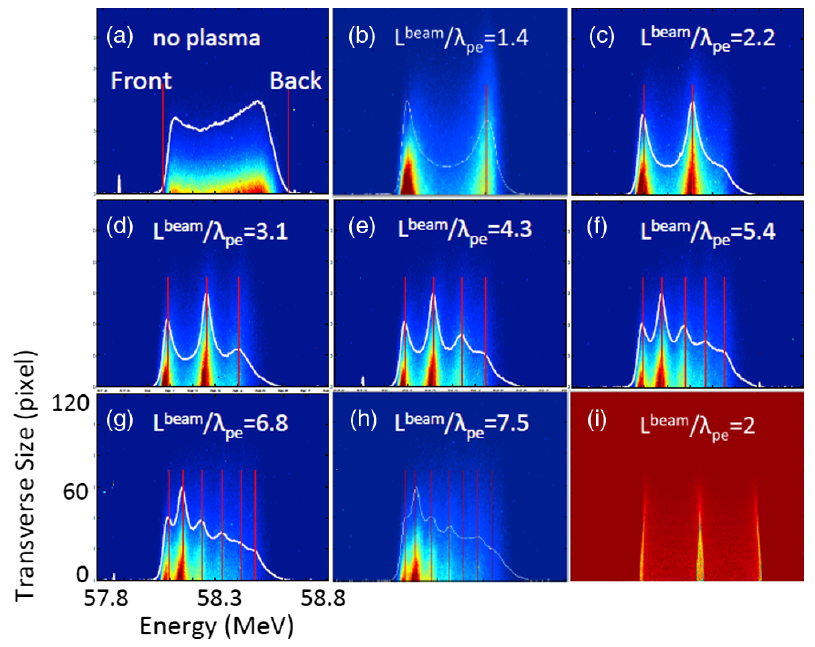}
\caption{ Energy spectra obtained at various plasma densities with a bunch with fixed length L$^{beam}$. %
Spectra (a) with n$_{e0}$=0 (no plasma) and (b)-(h) with increasing plasma densities between 2$\times$10$^{15}$ and 8$\times$10$^{16}$\,cm$^{-3 }$ obtained in the experiment. %
We note that in this case the incoming bunch density (a) is not constant along the bunch, but physics remains the same. %
The white lines indicate the sum of the images over the vertical image dimension (with no dispersion), and the red lines show the positions of the density peaks. %
Figure (i) shows a simulated energy spectrum for the case L$^{beam}$=\lpe/2, very similar to the experimental case of (c). %
In the experiment the plasma density needs not be set to integer values of L$^{beam}$=\lpe. %
The color tables are deliberately chosen different to avoid possible confusion between experimental and simulation results. %
Figure from Ref.~\cite{bib:fang}, with permission.}
\label{fig:FangMulti}
\end{figure}

Evidence that long particle bunches drive wakefields over multiple periods, as suggested for example by Eq.~(\ref{eq:MultiBunches_tran3}) was demonstrated experimentally~\cite{bib:fang}, as shown on Fig.~\ref{fig:FangMulti}. %
Figure~\ref{fig:FangMulti} shows a number of energy spectra measured with a bunch with fixed length L$^{beam}$ after propagation along a 2\,cm-long capillary plasma with various electron densities such that L$^{beam}/\lambda_{pe}$ varies essentially between one an~eight. %
The incoming bunch has an energy linearly correlated with time or longitudinal position along the bunch. %
This is known as an energy chirp. %
Therefore, measuring changes in energy as a function of energy is a~relative measure of the time structure of the longitudinal wakefields that were excited along the bunch. %
The~number of wakefields periods increasing with  L$^{beam}/\lambda_{pe}$ (Figs.~\ref{fig:FangMulti}(b)-(h)) is evident from the~spectra. %
The quantitative results are in excellent agreement with linear theory predictions and numerical simulation results (Fig.~\ref{fig:FangMulti}(i)). %
These wakefields could be used as seed for the self-modulation process described below. %
In fact, these results are direct evidence that the SM process can be seeded by a particle bunch with a sharp ($<$\lpe) rising edge in its current or density profile. %

\section{Bunch train}

We consider the case of bunches with square longitudinal density or current profiles and constant radius. %
In this case the results of 2D linear PWFA theory reduce to a 1D model where radial dependencies R(r) and dR(r)/dr in W$_z$ and W$_{\perp}$ (Eqs.~(\ref{eq:MultiBunches_long3}), (\ref{eq:MultiBunches_tran3}) and (\ref{eq:MultiBunches_long4}), (\ref{eq:MultiBunches_tran4})) can be considered as constant. %
For that to be valid we have to neglect transverse evolution upon propagation. %
Longitudinal evolution can generally be neglected because we consider propagation lengths over which relative dephasing between particles is small and can be neglected. %
That is, the following considerations are not valid any more once dephasing due to incoming or acquired energy differences must be included. %
With the above assumptions, Eqs.~(\ref{eq:MultiBunches_long1})~and~(\ref{eq:MultiBunches_tran2}) for wakefields within the bunch simplifies to Eqs.~(\ref{eq:MultiBunches_long3}) and~(\ref{eq:MultiBunches_tran3}). %
Here n$_b(\xi')$=n$_{b0}$=cst along $\xi'$ can be pulled out of the integral that can be simply calculated. %

Considering the bunch length $\xi_b$ one sees that the decelerating field within  the bunch is maximum for k$_{pe}\xi_b$=$\pi$/2 or $\xi_b=\lambda_{pe}$/4. %
The field behind the bunch reaches a maximum for $k_{pe}\xi_b=\pi$ or $\xi_b=\lambda_{pe}$/2. %
Indeed up to $\xi_b=\lambda_{pe}$/2 all particles lose energy. %
Making the bunch longer would mean adding particles that would find themselves in the accelerating phase of the wakefields and would gain energy from them and decrease their amplitude behind the bunch (as shown though the transformer ratio on Fig.~3 of~\cite{bib:muggliCAS}). %
We note that making the bunch 2$\pi$ longer yields the same wakefields amplitudes within and behind the bunch (periodic functions). %
We also note here that for k$_{pe}\xi=\pi$ the transformer ratio is two and is maximum. %
In this case the maximum possible energy extraction efficiency is limited by the fact that the decelerating field varies along the bunch to $\eta=\frac{1}{\pi}\int_0^{\pi}sin(x)dx\cong$64\%. %
The {\it first} and {\it last} particles do not lose any energy and the process continues until dephasing becomes an issue for particles losing energy at the highest rate (those with k$_{pe}\xi=\pi$ /2). %
We note that maximum efficiency with this sinusoidal variation of the decelerating field along the bunch reaches 73\% for \kpe$\xi_b\cong$0.73. %
The transverse wakefields expression shows that all particles till k$_{pe}\xi=\pi$ are focused (or at least not defocused) by the wakefields. %
For such a constant density bunch, the transverse wakefields are focusing all along the bunch: W$_{\perp}$ does not change sign along the bunch. %
The sign changes some distance behind the drive bunch. %
Since plasma electrons are expelled/attracted by a bunch with negative/positive charge sign, the inherent |${\bf E}$+${\bf v}_b\times{\bf B}$|$\rightarrow E/\gamma_b^2\rightarrow0$ for $\gamma_b\rightarrow\infty$ (or $\rightarrow$ small for $\gamma_b$ $\rightarrow$ large) force balance of relativistic beams is broken and the  ${\bf v}_b\times{\bf B}$ term focuses the bunch. %
We also remind the reader that behind the driver, the longitudinal and transverse wakefields oscillate around zero and are $\pi$/2 out of phase, at least in linear theory, as can be seen from Eqs.~(\ref{eq:MultiBunches_long4}) and (\ref{eq:MultiBunches_tran4}). %

\begin{figure}
\centering\includegraphics[width=.9\linewidth]{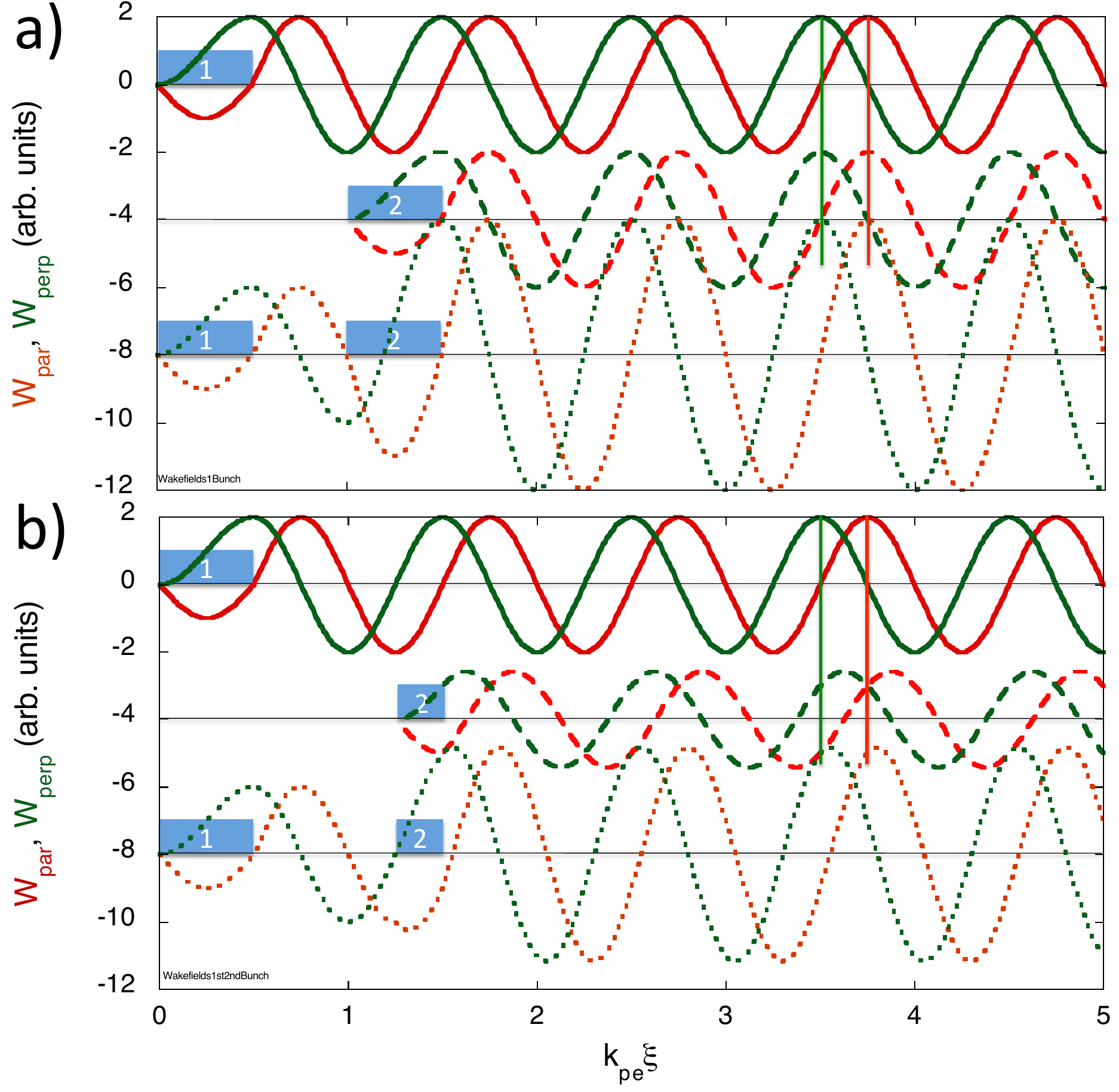}
\caption{Longitudinal (red lines) and transverse (green lines) wakefields driven by a single lead bunch (labelled 1), a following bunch (labelled 2, shifted by -4 units) and the train of two bunches (shifted by -8 units) as a function of position along the train $\xi$ normalized to the wavenumber \kpe (or phase normalized to 2$\pi$). %
a) Both bunches are \lpe/2-long and the second bunch is a distance \lpe (or phase 2$\pi$ behind the first one. %
b) The first bunch is \lpe/2-long, the second one \lpe/4-long and the second bunch is a distance $\frac{5}{4}\lambda_{pe}$ (or phase $\frac{5}{2}\pi$ behind the first one. %
Red and green vertical lines are aligned with one of the crests of the respective wakefields to allow for phase shift comparison. %
Bunches move to the left and wakefields are zero ahead of the first bunch.}
\label{fig:Wakefieldstrain}
\end{figure}

Considering a bunch train with the first bunch \lpe/2-long, it is natural to place the second bunch in the next (or some 2$\pi$ phase later) decelerating region of the wakefields to add energy to these fields. %
Two options for this bunch: make it also \lpe/2-long and in this case covering all the decelerating phase region, but also a focusing and a defocusing region, or rather \lpe/4-long and covering only the decelerating {\it and} focusing phase region. %
We show wakefields for both cases on Fig.~\ref{fig:Wakefieldstrain}. 
In both cases the wakefields behind the train are larger than behind the first one alone, as expected. %
They are larger in the~\lpe/2 than in the~\lpe/4-long-long case. %
While the second bunch \lpe/4-long is more likely to propagate because all in the focusing wakefields, transverse evolution (if considered) would also lead to wakefields evolution upon propagation along the plasma, unless a matching condition could be found (see~Ref.~\cite{bib:muggliCAS}). %
We note from Fig.~\ref{fig:Wakefieldstrain} that while the transformer ratio for the first bunch (alone) is two, as expected, it is less than two for the train (=4/3$\cong$1.33, see below, and $\cong$3.1/2.0=1.55 for the two bunch lengths, respectively). %
Since in the~first case the two bunches are identical and separated by 2$\pi$ in phase, the wakefields of individual bunches and of the train are in phase, as indicated by the vertical lines on the figures. %
However, in the~second case the 2$\pi$ symmetry is broken and the fields are not in phase with each other. %
Looking at fields behind the second bunch, in both cases the third bunch (and following ones) would follow the~second one with a 2$\pi$ phase shift. %
We also note that linear theory can be used to evaluate seed wakefields, for example as produced by a long bunch with a sharp rising edge or cut~\cite{bib:fang,bib:turner} or by a~relativistic ionization front traveling together and within the long bunch~\cite{bib:muggli}. %

From these simple examples we can use arithmetic to study the case of a train that would increase the transformer ratio rather than decrease it: the ramped bunch train~\cite{bib:rbt}. %
In this case all bunches are \lpe/2-long (because in this case we can do the arithmetic). %
The amplitude of the wakefields in each half period can be written as a (normalized) sequence (see first bunch Fig.~\ref{fig:Wakefieldstrain}(a)): -1:+2:-2:+2:-2:+2:$\ldots$ %
The second bunch, identical to the first one, placed in the~decelerating phase of the first one drives relative amplitudes: 0:0:-1:+2:-2:+2:$\ldots$
The sum of the amplitudes is therefore: -1:+2:-3:+4:-4:+4:$\ldots$ %
The transformer ratio is thus indeed 4/3$\cong$1.33, as seen on Fig.~\ref{fig:Wakefieldstrain}(a), smaller than two, because the (total) decelerating field within the second bunch is larger than the within the first bunch. %
One can make the~decelerating field in the second bunch equal to that of the first one by making its density (with its charge) three times larger than that of the~first one and placing it into the accelerating phase of the~first one (a +2 amplitude region). %
It then drives a sequence: 0:0:0:-3:+6:-6:$\ldots$ %
The wakefields amplitudes behind the two bunches then become (the sum): -1:+2:-2:-1:+4:-4:$\ldots$%
The transformer ratio is now equal to four. %

Let us understand what is happening. %
The first bunch loses energy to the wakefields at rate -1. %
The~second one by itself would lose energy at a rate -3, but gains from the first one at rate +2. %
Net energy is therefore transferred from the first to the second bunch. %
A witness particle (or bunch with negligible charge, not beam loading, charge $\ll$1 in these normalized units) placed at the appropriate phase would lose energy to the wakefield of the first bunch at rate -2, but gain from the wakefields of the second bunch at rate +6. %
As mentioned above, we neglect it wakefields due to its very small relative charge. %
Thus the~total gain rate is +4. %
The sequence of drive bunches can be continued, all placed in the~next acceleration phase (2$\pi$ phase distance) with relative charges 1:3:5:7:$\ldots$ reaching transformer ratios 2,4,6,8,$\ldots$ %
In this scheme, energy is transferred from the preceding to the next bunch to guarantee its -1 energy loss rate for all bunches. %
At the end, all bunches can lose all their energy (till dephasing becomes an issue), the same way a single bunch would. %

One can compare the case of two bunches with relative charge one and three to that of a single one with charge four. %
In the train case, the loss rate E$_-$ is -1 and the gain rate for witness particles +4. %
Energy depletion occurs over some length L (L=W$_0$/E$_-$) and the energy gain per witness particle can be four times the incoming particles energy W$_0$. %
In the single bunch case (with the same total normalized charge of 4), the loss rate is -4 and the gain rate +8. %
Energy depletion occurs over a length four times shorter than in the train case, but particles can gain only twice the incoming particles energy (assumed to be the same in both cases for comparison). %
Since the total energy must be conserved, in the first case four times less particles can gain energy (same number of incoming particles with the same energy in both cases). %
Energy transfer efficiencies are the same assuming all the energy lost by drive particles is gained by witness ones. %
Using a single bunch or a train of bunches is therefore a matter of application and opportunity. %
However, we note as an example that by shaping a single bunch or by using a bunch train, 5\,GeV electrons could be produced out of 1\,GeV electrons, for example for FEL applications for which a low charge (but high current) witness bunch (with small energy spread and emittance) may be sufficient (see for example Ref.~\cite{bib:SPARC}). %

We note here that the ramped bunch train needs not to have gaps between bunches. %
One can imagine a staircase train with charge 1:3:5:7:$\ldots$ in each \lpe/2 interval, a crude approximation of the~triangular single bunch shape that was proposed in a broader wakefields context~\cite{bib:bane}. %
This bunch drives decelerating fields that oscillate with the sin($k_{pe}\xi$) and leads to a poor energy transfer efficiency (on the oder of 64\% as noted above). %
However, adding a preceding doorstep (constant charge bunch over a $\lambda_{pe}$/4 length) or other preceding features can greatly enhance the overall efficiency. %

We note again that bunches and wakefields evolution upon propagation and under the influence of transverse wakefields probably makes the bunch train solution in the linear regime impractical for large energy gain (because of long propagation distance). %
A train with the same purpose (large transformer ratio) can be built for the nonlinear regime where self-similar propagation may be possible. %
In this case (weak) plasma electron blow-out in what is sometimes coined as quasi-linear or weakly non-linear PWFA regimes may mitigate the some of the effects of bunch transverse evolution. %
However, no recipe (as presented above) exists for the bunch parameters. %
One must resort to numerical simulations to design the system~\cite{bib:fangtrain}. %

In this context, one may look for a way for a long ($\gg$\lpe), continuous bunch to produce the bunch train through its interaction with wakefields: self-modulation. %

\section{Self-modulation}
Self-modulation was first discussed for long laser pulses~\cite{bib:SMlaser}. %
Equations describing the SM of laser pulses and those of particle bunches are similar, although the physics at play is quite different. %
There are a number of theoretical and simulation papers describing in some details the SM of long particle bunches~\cite{bib:kumar,bib:pukhov,bib:lotov}. %
Here we use again simple arguments to give a feel for the process. %

We consider again a long, relativistic, uniform density and radius charged particle bunch of length L$\gg\lambda_{pe}$ with density smaller than the plasma electron density. %
At the beginning of the plasma (z=0), the transverse wakefields are only focusing (or null) for bunch particles (see Eq. (~\ref{eq:MultiBunches_tran3})). %
These wakefields are called seed wakefields in the context of development and control of the SM instability. %
Upon a short propagation distance, the effect of these wakefields is to change the bunch (RMS) radius and thus to modulate its density. %
The ever so slightly modulated bunch drives wakefields that are different from those driven by the incoming, un-modulated bunch for two reasons. %
First, a modulation of the bunch radius enters in the wakefields calculations through the bunch radial dependencies R(r) and dR/dr. %
In~what follows we again neglect this effect. 
Second, a change in radius changes the bunch density. %
One expect the density to increase according to the initial transverse wakefields. %
Wakefields are harmonic, one may thus expect the change in bunch radius to have the same dependency as the wakefields, i.e. write $\sigma_r=\sigma_{r0}-\epsilon\left(1-cos(k_{pe}\xi)\right)$. %
Since the bunch density is proportional to the inverse of the bunch radius square (n$_b$($\xi)\propto$1/$\sigma_r^2(\xi)$), one can assume:
\begin{equation}\label{eq:nbev}
n_b\propto\frac{1}{\sigma_r^2}\cong\frac{1}{\sigma_{r0}^2}\left(1+2\epsilon(1-cos(k_{pe}\xi))\right)\propto n_{b0}\left(1+2\epsilon(1-cos(k_{pe}\xi))\right). 
\end{equation}
Here, $\epsilon\ll1$ is a small parameters reflecting the small variation in bunch density resulting from the small variation in radius over a small propagation distance of the bunch into the plasma. %
In this expression one recognizes the un-modulated density of the bunch, $\propto n_{b0}\left(1+2\epsilon\right)$, resulting from the adiabatic response of the plasma that focusses the entire bunch, and a newly created density modulation: $-2\epsilon cos(k_{pe}\xi)$. %
The~first two terms (without modulation) yield wakefields of the same form, but with slightly larger amplitude than seed wakefields since the adiabatic response of the plasma slightly increases the overall bunch density. %
The third term can be inserted in the linear wakefield equation~(\ref{eq:MultiBunches_tran1}) and the equation integrated\footnote{We used: sin($\alpha\pm\beta$)=sin($\alpha$)cos($\beta$)$\pm$cos($\alpha$)sin($\beta$), cos($\alpha\pm\beta$)=cos($\alpha$)cos($\beta$)$\mp$sin($\alpha$)sin($\beta$) and $\int$cos$^2$($\alpha)$d$\alpha$=$\frac{\alpha}{2}$+$\frac{sin(2\alpha)}{4}$+C, $\int$sin($\alpha$)cos($\alpha$)d$\alpha$=-$\frac{1}{4}$cos(2$\alpha$)+C} to yield an additional term to the wakefields:
 \begin{equation}\label{eq:modwakef}
\delta W_{\perp}\propto-\epsilon\xi sin(k_{pe}\xi). 
\end{equation}
This (small) term has an amplitude that is proportional to $\xi$ and corresponds to wakefields increasing along the bunch. %
This is an effect analogous to that of a bunch train (a sort of modulation) driving wakefields, as described here above. %
In addition, this term is also harmonic, but with a phase different from that of the seed wakefields. %
The sum of these with the seed wakefields therefore also has a different phase than the initial wakefields. %
Since the first modulation maximum is located at $\xi$=\lpe/2 back from the bunch beginning, the sum wakefields phase is shifted backwards with respect to the seed wakefields. %

This simple evaluation shows some important characteristics of the SM process development: Wakefields grow along the plasma, i.e., have larger amplitude even after a small propagation distance into the plasma; Wakefields grow along the bunch ($\xi$ dependency in Eq.~(\ref{eq:modwakef})); Wakefields phase shifts backward along the bunch during growth, therefore wakefields phase velocity is slower than that of the bunch~\cite{bib:pukhov}. %
We note that all these points were of course predicted by theory~\cite{bib:kumar} and the first two were so far demonstrated experimentally~\cite{bib:turner}. %

Further development of the SM process must be performed with full versions of the beam envelope equation~\cite{bib:schroederSM} or through numerical simulations. %
One of the interesting questions is whether the SM process converges towards a final and stable situation similar to that used for Fig.~\ref{fig:Wakefieldstrain}(b): all bunch particles reside only in the decelerating and accelerating phase of the wakefields. %
Also, what fraction of the drive particles remain in the drive bunches once a bunch train configuration that can propagate self-similarly over long plasma length has been reached? %
Ignoring the evolution of the particles distribution due to the~change in phase velocity of the wakefields~\cite{bib:pukhov}, we note that the SM process is governed by transverse wakefields, and not longitudinal ones. %
In Section~\ref{sec:lintheo} we used only longitudinal wakefields considerations to design the bunch train. %
One can imagine then that particles in the focusing phase of the wakefields remain. %
However, in linear wakefields, the focusing phase covers accelerating and decelerating phases. %
If those were equally populated, little wakefields would be left behind each new-formed bunch since energy lost by the first half the particles would be absorbed by the other half. %
Simulation and experimental results show that the evolution leads to a situation somewhere in between~\cite{bib:karl,bib:turner}. %

Regarding the transformer ratio that SM may produce, we showed above that, in the case of a~bunch train, producing R$>$2 requires bunch shaping with particular density ratio (through charge at constant radius). %
Proton synchrotrons or electron linacs usually produce (long) bunches with Gaussian or quasi-Gaussian current or density profile. %
The SM process mostly retains this incoming profile and even if self-consistent evolution may alter that profile, it is unlikely that it would create the proper profile for reaching large R value wakefields. %
We can therefore expect that SM will produce R=1 wakefields (limit with many similar density bunches). %
Shaping of the incoming bunch profile (before SM, if possible) may be used to lead to wakefields with R$>$1. %

\begin{figure}
\centering\includegraphics[width=.9\linewidth]{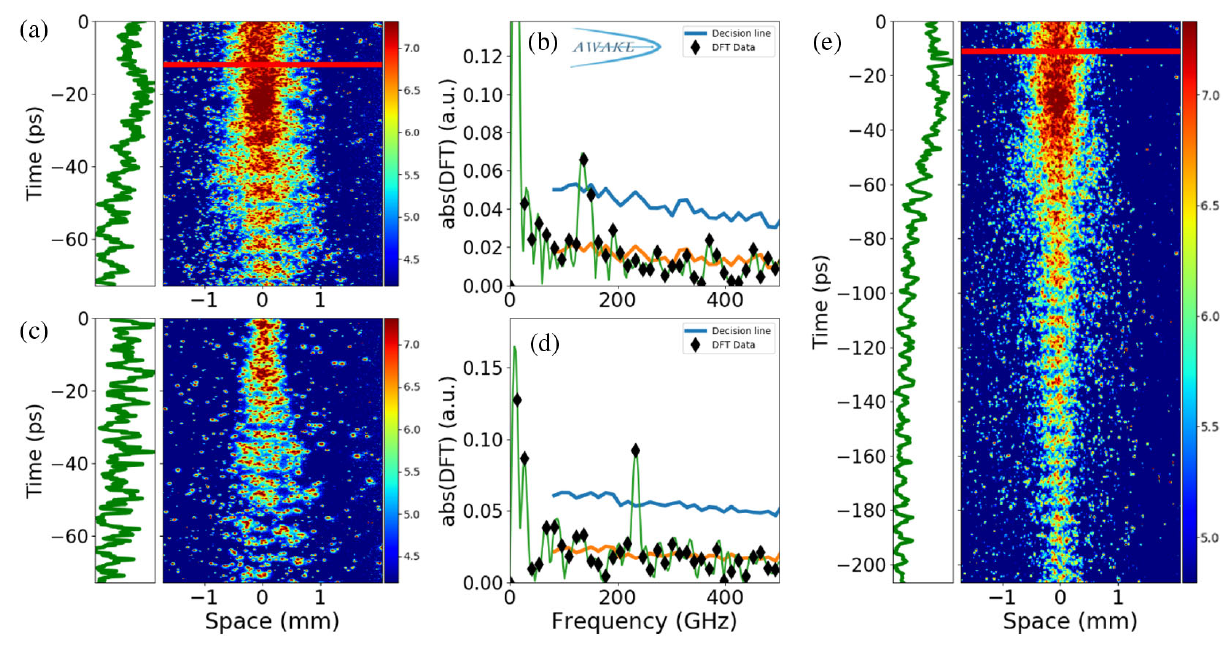}
\caption{Streak camera images (a) at low (n$_{Rb}$=2.457$\times$10$^{14}$\,cm$^{-3}$) and (c) at high
(n$_{Rb}$=6.994$\times$10$^{14}$\,cm$^{-3}$) plasma densities. Profiles obtained by summing the images along the spatial axis from -0.4 to 0.6\,mm are displayed on the left-hand side of each image. The profile of image (a) shows the defocusing effect of SSM starting at the laser pulse time ($\sim$10\,ps). Image (c) is obtained $\sim$10\,ps behind the ionizing laser pulse that is placed in the~middle of the bunch. %
Figures (b) and (d) show the DFT power spectrum for the two profiles (black diamonds) as well as for background images (orange lines). %
The green lines depict the interpolated power spectrum. %
The blue lines show a noise threshold used for automatically detecting frequencies. Image (e) shows a low density case (n$_{Rb}$=2.190$\times$10$^{14}$\,cm$^{-3}$) where the full train of micro-bunches is shown. %
The Rubidium (and thus plasma) density for image (e) has an upwards density gradient of 0.34\%/m. %
Figure from Ref.~\cite{bib:karl}, with permission.}
\label{fig:SMresults}
\end{figure}

Self-modulation of a long proton bunch was demonstrated experimentally~\cite{bib:karl}. %
Figure~\ref{fig:SMresults} shows time-resolved images of the proton after 10\,m of plasma at various densities. %
The presence of micro-bunches in the back of the proton bunch is evident from the pictures (a), (c) and (e). %
Fourier analysis (b) and (d) shows that the bunch modulation frequency is the plasma frequency, f$_{pe}$=\ope/2$\pi$. %
 
\section{Curent filamentation instability}

\subsection{Plasma return current}

As was noted in the plasma lectures, the response of plasma electrons (plasma ions remain essentially immobile at the 1/$\omega_{pe}$ time scale) tends to cancel fields of an external charge. %
For the cold plasmas we consider and for time-varying perturbations leading to wakefields excitation, cancellation occurs at the~1/\ope~time, and c/\ope~spatial scales. %
When a relativistic bunch of particles of charge q$_b$ with density n$_b$ and velocity v$_b\cong c$ enters the plasma it represents a current density j$_b$=q$_b$n$_b$v$_b$. %
This current density carries an azimuthal (cylindrical symmetry assumed) magnetic field that can be deduced from Maxwell equation: $\nabla\times B=\mu_0j_b$. %
At any location along the plasma the transverse magnetic field is zero before, maximum during, and zero again after the bunch passage. %
This magnetic field varying in time generates a~flus variation through a small (imaginary) transverse loop, which generates a varying electric field around the loop according to $\nabla\times E=-\partial B/\partial t$. %
Under the influence of this electric field, plasma electrons generate a plasma current that tends to cancel the bunch-generated flux, current that is then opposed to that of the bunch. %
This current is called the {\it plasma return current}. %
It is interesting to note that in the case of a negatively charged bunch, bunch particles and plasma electrons have opposite velocities, whereas in the case of a positively charged bunch, bunch particles and plasma electrons have velocities in the same direction. %
We also note that in the context of linear regime for which n$_b\ll n_{e0}$, the~plasma electron (longitudinal in this case) velocity is proportionally smaller than the bunch particles one: $j_b=q_bv_bn_b\cong en_{e0}v_e=j_e$, thus $v_e\cong\frac{n_b}{n_{e0}}v_b\ll c$. %
Transversely, the response of the plasma electrons is again at the~c/\ope~scale. %
Therefore, in the case of a narrow bunch, meaning such that \kpe$\sigma_r\le$1, most or all the~plasma return current flows {\it outside} the~bunch. %
The bunch is therefore not affected by the return current. %
However, in the opposite case of a wide bunch (\kpe$\sigma_r>$1), the return current flows within the~bunch. %
The effect of plasma return current was seen for example in plasma lens experiments~\cite{bib:plasmalens}. %

\subsection{Current filamentation instability}

In the wide bunch case ( \kpe$\sigma_r\gg$1), two opposite direction currents flow through each other. %
When exactly compensating for each other, the net current is zero and the situation is stable (no net force). %
However, when exact compensation does not occur, the two locally un-equal currents repel each other. %
The stronger current repels the weaker one and self-focuses, increasing its current density and azimuthal magnetic field, yielding a positive feedback mechanism. %
The transversely large bunch transforms into a~series of narrower ones, carrying larger current density and magnetic field. %
Each of them has a transverse size smaller than c/$\omega_{pe}$. %
This instability is called the current filamentation instability (CFI)~\cite{bib:CFI}. %
It~can be seen as the relativistic version of the Weibel instability~\cite{bib:weibel}. %

\begin{figure}[h]
\centering\includegraphics[width=.9\linewidth]{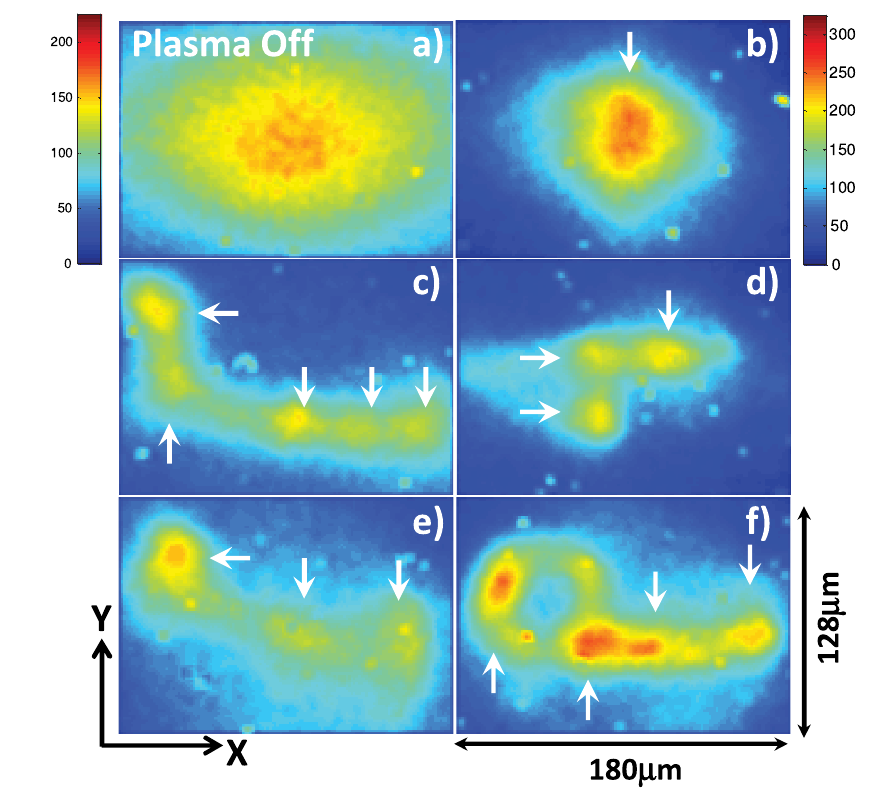}
\caption{Images of the bunch at the capillary exit with arrows indicating filaments. %
Bunch parameters are: $\sigma_{x0}$=80\,$\mu$m, $\sigma_{y0}$=50\,$\mu$m, and charge Q$\cong$1.0\,nC. (a) Without plasma and (b) n$_{e0}$=1.6$\times$10$^{16}$\,cm$^{-3}$ ($k_{pe}\sigma_{y0}$=1.2), (c) and (d) n$_{e0}$=1.2$\times$10$^{17}$\,cm$^{-3}$ ($k_{pe}\sigma_{y0}$=3.3) and (e) and (f ) n$_{e0}$=1.9$\times$10$^{17}$\,cm$^{-3}$ ($k_{pe}\sigma_{y0}$=4.1). X-rays are visible in the images and are a few pixels in size. %
Color tables are the same for (b) through (f). %
Figure from Ref.~\cite{bib:CFIexp}, with permission.}
\label{fig:CFIExp}
\end{figure}

\begin{figure}[h]
\centering\includegraphics[width=.9\linewidth]{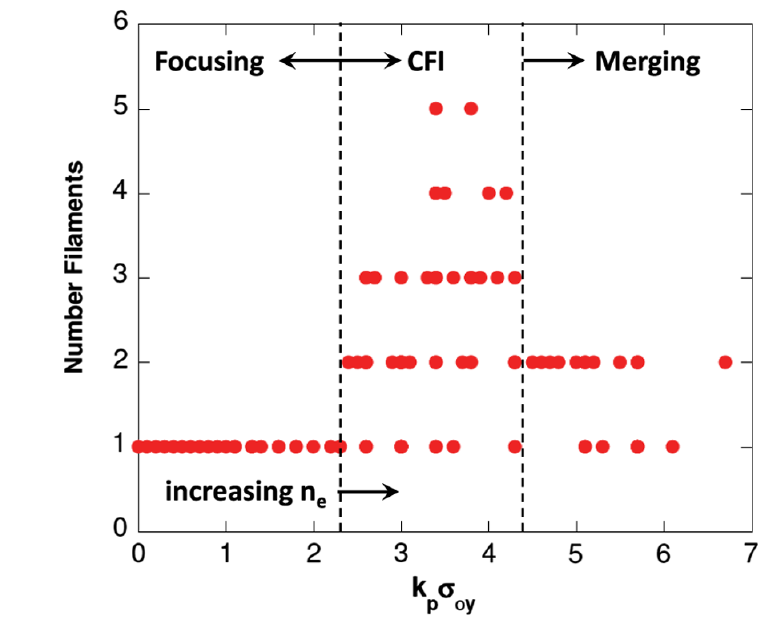}
\caption{ Number of filaments observed in single events as a function of the CFI parameter $k_{pe}\sigma_{y0}$. For this measurement the charge Q$\cong$1.0\,nC, $\sigma_{x0}$=80\,$\mu$m, $\sigma_{y0}$=50\,$\mu$m. %
Similar events lead to only single filaments for $k_{pe}\sigma_{y0}<$2.2 and multiple (one to five) filaments for $k_{pe}\sigma_{y0}<$2.2. For $k_{pe}\sigma_{y0}>$4.5 only one and two filaments are seen and could be due to merging of CFI generated filaments. %
In general single points represent multiple events at the same $k_{pe}\sigma_{y0}$. %
Figure from Ref.~\cite{bib:CFIexp}, with permission.}
\label{fig:CFITrans}
\end{figure}

Current filamentation instability was observed with an electron bunch and a threshold for instability at $k_{pe}\sigma_r\cong$2.2~\cite{bib:CFIexp}. %
An example of observation is shown on Fig.~\ref{fig:CFIExp}. %
This figure~\cite{bib:CFIexp} shows a number of bunch transverse profiles as measured at the end of a 2\,cm-long capillary plasma for various plasma densities. %
These correspond to various values of $k_{pe}\sigma_r$ obtained by changing the plasma density for a~constant $\sigma_r$. %
Evolution of the transversely Gaussian incoming bunch (no plasma, Fig.~\ref{fig:CFIExp}(a) and plasma lens focusing, Fig.~\ref{fig:CFIExp}(b)) into multiple filaments is evident from the images. %
Figure~\ref{fig:CFITrans} shows the number of filaments observed as a function of $k_{pe}\sigma_r$ ($k_{pe}\sigma_{y0}$ in this case). %
Transition from focusing for small $k_{pe}\sigma_r$, to multiple filaments around $k_{pe}\sigma_r\cong2.2$ is evident. %
Merging of filaments at large $k_{pe}\sigma_r$ values ($>$4.5) is also observed. %

It is clear that CFI has to be avoided in the PWFA context. %
The condition $k_{pe}\sigma_r=1$ is used to determine the maximum plasma density at which an experiment with a beam focused to a minimum transverse size $\sigma_r$ can operate~\cite{bib:muggli}. %
In astrophysical context, CFI or Weibel instabilities of plasmas flowing into the (plasma) inter-stellar medium may be responsible for generation of large and small scale magnetic fields. %
In particular, instabilities evolution may explain some of the spectral features observed in the emission of x- and $\gamma$-rays detected on Earth~\cite{bib:jitter}, possibly due to what is known as {\it jitter radiation}. %
Another interesting effect is the instability of neutral flowing plasmas, sometimes referred to as {\it fire balls}, flowing in through the interstellar medium. %
These {\it fire balls} could be studied in experiments by using neutral beams, {\it fire ball beams}~\cite{bib:fireball}, composed of equal number of electrons and positrons. %

\section{Final remarks}
We attempted to give simple, intuitive and {\it physical} descriptions of self-modulation and transverse filamentation of a long or wide relativistic particle bunch propagating in a plasma~\cite{bib:muggli3}. %
The descriptions have evident limitations and use many hypotheses and short-cuts. %
The reader is thus advised to consult published papers on the topics that present more rigorous descriptions of the processes described here. %
Many further details can be found in the references chosen here and in the many others out there. %
We also showed sample experimental results that demonstrate that the physics at play has, to some extent, been observed experimentally. %
There is also a large body of numerical simulation results (not mentioned here) that largely support and detail theory results. %
The text is thus an introduction meant to encourage the~novice reader to further research all these topics and to build his/her own understanding of the~physics at play. %
Plasma wakefield acceleration is an interesting and challenging topic that has experimentally matured over the last two decades. %
Building an accelerator on its principle is a challenge that requires understanding of many topics, and maybe most of all, the understanding of the concept of compromise. %

\end{document}